\documentclass[%
 aip,
 sd,%
 amsmath,amssymb,
 reprint,%
]{revtex4-1}

\usepackage{graphicx}
\usepackage{dcolumn}
\usepackage{bm}

\begin{document}

\preprint{AIP/123-QED}

\title[Time-dependence of the transmission matrix of a specialty few-mode fiber]{Time-dependence of the transmission matrix of a specialty few-mode fiber}

\author{J. Yammine}
\affiliation{%
Univ. Lille, CNRS, UMR 8523 - PhLAM - Physique des Lasers Atomes et Mol\'{e}cules, F-59000 Lille, France.
}%
\author{A. Tandj\`{e}}
\affiliation{%
Univ. Lille, CNRS, UMR 8523 - PhLAM - Physique des Lasers Atomes et Mol\'{e}cules, F-59000 Lille, France.
}%
\affiliation{%
\'{E}cole Polytechnique d'Abomey-Calavi (EPAC), Univ. d'Abomey-Calavi (UAC), Cotonou, B\'{e}nin
}%
\author{Michel Dossou}
\affiliation{%
\'{E}cole Polytechnique d'Abomey-Calavi (EPAC), Univ. d'Abomey-Calavi (UAC), Cotonou, B\'{e}nin
}%

\author{L. Bigot}
\affiliation{%
Univ. Lille, CNRS, UMR 8523 - PhLAM - Physique des Lasers Atomes et Mol\'{e}cules, F-59000 Lille, France.
}%
\author{E. R. Andresen}
\email{esben.andresen@univ-lille.fr}
\affiliation{%
Univ. Lille, CNRS, UMR 8523 - PhLAM - Physique des Lasers Atomes et Mol\'{e}cules, F-59000 Lille, France.
}%

\date{\today}

\begin{abstract}
We report a time-resolved measurement of the full transmission matrix (TM) of a short length of specialty annular-core few-mode fiber which guides 10 vector modes. We show how our method can isolate the fiber TM from "misalignment" contributions from optics upstream and downstream of the fiber. From measurements spanning two days we extract the drift of the fiber TM. We show that drifts in the TM elements are mostly described as correlated phase variations rather than amplitude variations. We show that an empirical model of the fiber TM parametrized in one parameter can successfully account for the drift. 
\end{abstract}

\maketitle


\section{\label{sec:Intro}Introduction}

Currently, the utilized capacity of the World's single-mode fiber (SMF)-based communication networks is nearing their physical limits \cite{MitraNature2001}. Spatial division multiplexing (SDM) based on multi-core fiber (MCF) or mode division multiplexing (MDM) based on few-mode fiber (FMF) have the potential for increased capacity.  \cite{WinzerNatPhoton2014, LiAOP2014, RichardsonNatPhoton2013}.  \\
In the simplest approximation, a MCF with $N$ cores is thought of as $N$ independent parallel channels, each being the equivalent of a SMF. Apart from very short links, this approximation is insufficient as the cores in general couple among each other leading to energy transfer between cores or inter-core cross-talk. 
The interaction of two cores is governed by coupled-mode theory \cite{SnyderJOSA1972}, so the cross-talk is a function of, among other, the effective index differences between the cores which, in turn, are quite sensitive to strain, temperature, etc. which must be expected to vary both as a function of distance and time. 
The time evolution of transmission quality in MCF has been the subject of several papers \cite{MachoOE2015, LuisJLT2016, RademacherECOC2017, PuttnamECOC2017, AlvesOE2017, AlvesJLT2017, AlvesECOC2017, GanOE2018, AlvesOE2018}. In effect, experimental measurements show that inter-core cross-talk can vary over time by as much as 10~dB \cite{MachoOE2015, LuisJLT2016, RademacherECOC2017, PuttnamECOC2017, AlvesOE2017, AlvesECOC2017}. The consequences for the design of MCF-based networks are therefore important---the variability of the inter-core cross-talk must be known in order to design a MCF network that can assure a certain minimum performance in all conditions. Channel models for MCF have been proposed in order to provide tools to assist in designing MCF networks \cite{AlvesJLT2017,GanOE2018, AlvesOE2018}. \\
The very same issues are expected to be present in FMFs. Nevertheless, to the best of our knowledge FMFs have not yet been subjected to the same level of scrutiny as their MCF counterparts in what concerns their temporal behavior. Indeed, we are not aware of time-resolved studies of the quality of transmission in FMF. 
Two questions concerning the time-behavior of FMF are essential: (i) how does the cross-talk between degenerate modes within a mode family evolve in time, a mode family being a grouping of modes whose effective indices are very close; (ii) how does the cross-talk between any two mode families evolve in time. These questions are not straightforward to address because we do not have easy access to the individual spatial channels (modes) of a FMF (as opposed to MCF where the individual spatial channels are the modes localized on the cores). FMFs are usually implemented interfaced with modal multiplexers or demultiplexers which in themselves can present non-negligible cross-talk, either intrinsic, or as a result of misalignment, making it difficult to isolate the cross-talk of the FMF alone. \\
Here, we present a method for isolating the transmission matrix (TM) of the FMF---the "fiber TM"---from the interfacing components. We use this to perform time-resolved measurements of the TM of a specialty, annular-core fiber supporting 10 vector modes and thus evaluate the temporal behavior of the multi-mode channel represented by the FMF.

\section{\label{sec:Experimental}Experimental}
\subsection{\label{subsec:fiber} Annular-core fiber}
\begin{figure}
\centering
\includegraphics[width=\columnwidth]{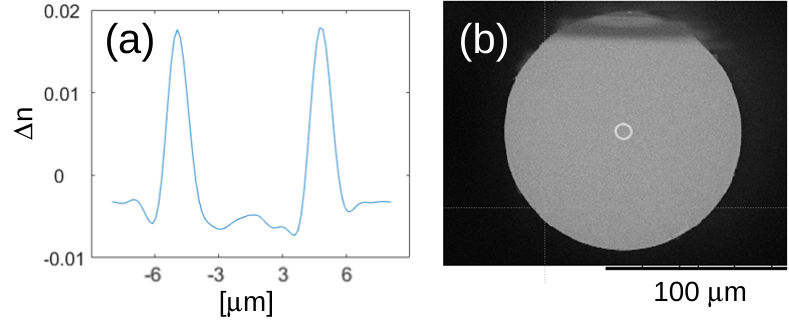}
\caption{\label{fig:fiber}
(a) Measured refractive index profile of the annular-core fiber as a function of the radial coordinate. (b) Scanning electron micrograph of the fiber. 
}
\end{figure}
\begin{table}[htbp]
\caption{\label{tab:fiber}The vector modes of the annular-core fiber and their calculated effective indices. }
\begin{displaymath}
\begin{array}{|c|c|}
  \hline
  \mathrm{Mode} & n_{\mathrm{eff}} \\
  \hline
  \hline
  \mathrm{HE}_{11}^{(e)}, \mathrm{HE}_{11}^{(o)} & 1.4484829 \\  
  \hline
  \hline
  \mathrm{TE}_{01} & 1.4471994 \\
  \hline
  \mathrm{HE}_{21}^{(e)}, \mathrm{HE}_{21}^{(o)} & 1.4471433 \\
  \hline
  \mathrm{TM}_{01} & 1.4470770 \\
  \hline
  \hline
  \mathrm{HE}_{31}^{(e)}, \mathrm{HE}_{31}^{(o)} & 1.4442856 \\
  \hline
  \mathrm{EH}_{11}^{(e)}, \mathrm{EH}_{11}^{(o)} & 1.4442890 \\
  \hline
\end{array}
\end{displaymath}
\caption{\label{tab:LP}The linearly polarized, helical-phase mode basis.}
\begin{displaymath}
\begin{array}{|c|c|c|}
  \hline
  \# &  (l,\mathrm{pol}) & \mathrm{Constituent~vector~modes} \\
  \hline
  \hline
  1 & (0,h) & \mathrm{HE}_{11}^{(e)} \\  
  \hline
  2 & (0,v) & \mathrm{HE}_{11}^{(o)} \\  
  \hline
  \hline
  3 & (-1,h) & (\mathrm{TE}_{01} - j \mathrm{TM}_{01}) + (\mathrm{HE}_{21}^{(e)} - j \mathrm{HE}_{21}^{(o)}) \\
  \hline
  4 & (1,h) &  (\mathrm{HE}_{21}^{(e)} + j \mathrm{HE}_{21}^{(o)}) + (\mathrm{TE}_{01} + j \mathrm{TM}_{01})\\
  \hline
  5 & (1,v) &  (\mathrm{HE}_{21}^{(e)} + j \mathrm{HE}_{21}^{(o)}) - (\mathrm{TE}_{01} + j \mathrm{TM}_{01})\\
  \hline
  6 & (-1,v) &  (\mathrm{TE}_{01} - j \mathrm{TM}_{01}) - (\mathrm{HE}_{21}^{(e)} - j \mathrm{HE}_{21}^{(o)})\\
  \hline
  \hline
  7 & (-2,h) &  (\mathrm{EH}_{11}^{(e)} - j \mathrm{EH}_{11}^{(o)}) + (\mathrm{HE}_{31}^{(e)} - j \mathrm{HE}_{31}^{(o)}) \\
  \hline
  8 & (2,h) &  (\mathrm{HE}_{31}^{(e)} + j \mathrm{HE}_{31}^{(o)}) + (\mathrm{EH}_{11}^{(e)} + j \mathrm{EH}_{11}^{(o)})\\
  \hline
  9 & (2,v) & (\mathrm{HE}_{31}^{(e)} + j \mathrm{HE}_{31}^{(o)}) - (\mathrm{EH}_{11}^{(e)} + j \mathrm{EH}_{11}^{(o)}) \\
  \hline
  10 & (-2,v) & (\mathrm{EH}_{11}^{(e)} - j \mathrm{EH}_{11}^{(o)}) - (\mathrm{HE}_{31}^{(e)} - j \mathrm{HE}_{31}^{(o)}) \\
  \hline
\end{array}
\end{displaymath}
\end{table}
For this study we used an annular-core fiber fabricated at the FiberTech Lille technology platform and whose measured index profile is shown in Fig.~\ref{fig:fiber}(a). A scanning electron micrograph of the same fiber is presented in Fig.~\ref{fig:fiber}(b). Its length was 1.5~m. 
We modelled the fiber using the fitted refractive index curve and predicted the presence of 10 guided vector modes which are listed in Tab.~\ref{tab:fiber} along with their calculated effective indices. From the calculated vector mode fields we construct a basis of linearly polarized modes with helical phase cf Tab.~\ref{tab:LP} which will be employed in the fiber TM measurement as will be described later. The modes are denoted as ($l$, pol) where $l$ is the topological charge or the number of 2$\pi$ phase turns per period and \textit{pol} is the polarization $v$ or $h$. Our choice of mode basis was motivated by the fact that the measurement was done in this basis and does not affect the result. Indeed, the measured fiber TMs can be transformed into any basis, and we refer to App.~\ref{app:altbasis} where we present measured fiber TMs in other bases.
This fiber is roughly similar to the fiber that was used in Ref.~\citenum{BozinovicScience2013} to demonstrate Terabit-scale orbital angular momentum mode division multiplexing. 

\subsection{\label{subsec:TM}Transmission matrix measurement}
\begin{figure*}[htbp]
\centering
\includegraphics{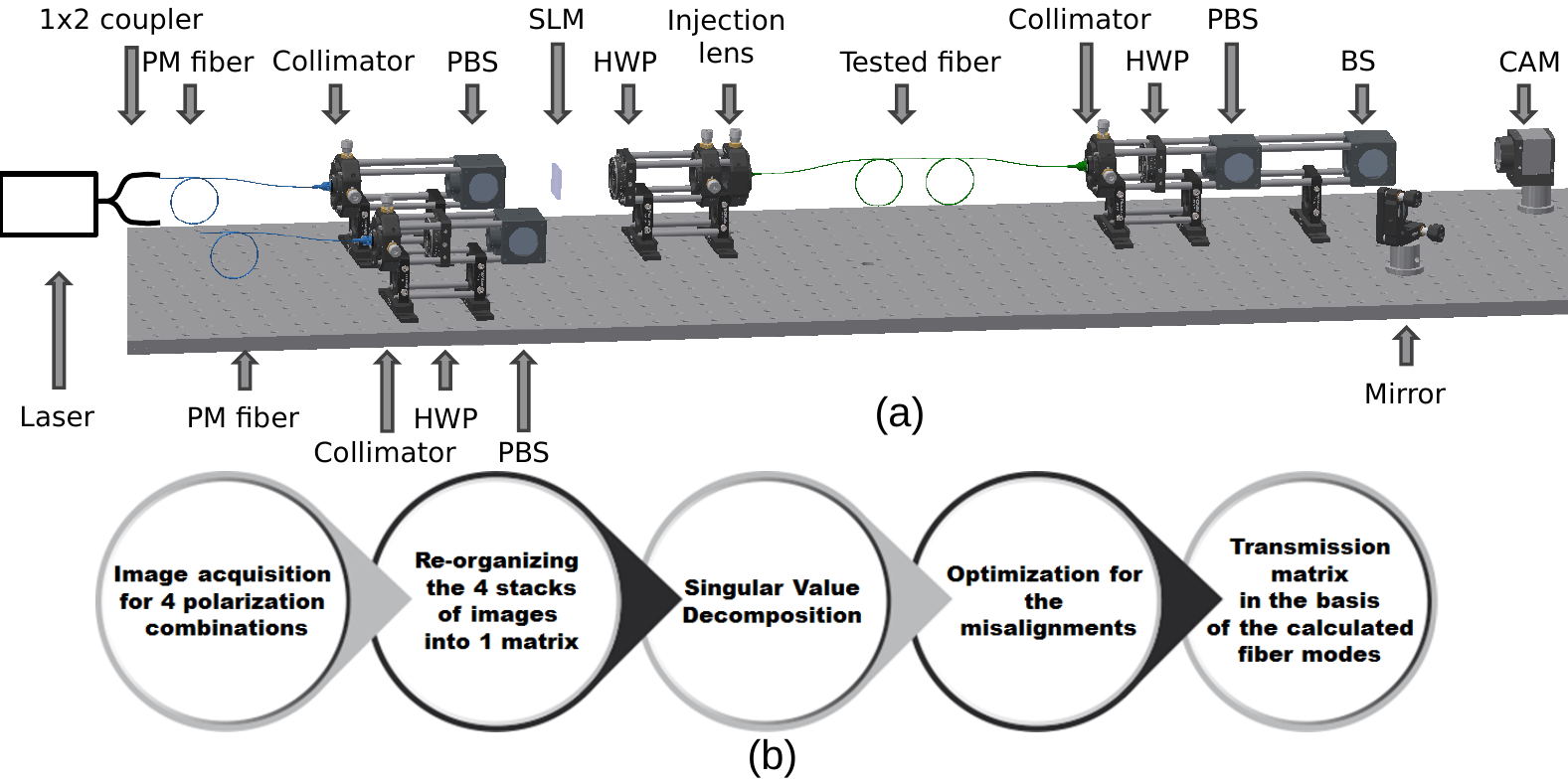}
\caption{\label{fig:TMmeas}
(a) Experimental setup. PM, polarization-maintaining; PBS, polarizing beam splitter; SLM, two-dimensional spatial light modulator; HWP, half-wave plate; BS, non-polarizing beam splitter; CAM, camera. 
(b) Flowchart of the fiber TM measurement.
}
\end{figure*}
The fiber TM is the matrix that links electric field going into every possible fiber mode to the electric field going out of every possible fiber mode. As such, the complete three-dimensional fiber TM with dimensions input mode, output mode, and wavelength completely describes the linear behavior of the fiber. In the present study we limit ourselves to one single wavelength, thus giving a fiber TM that contains the entire linear behavior of the fiber at one wavelength. The fiber TM measurement methodology used here is similar to ones employed in Refs.~\citenum{PloschnerNatPhoton2015} and \citenum{CarpenterNatPhoton2015}. For a detailed description, we refer to App.~\ref{app:TM}. \\
The setup is presented in Fig.~\ref{fig:TMmeas}(a). Thanks to a two-dimensional spatial light modulator (SLM) we can inject any localized input mode into the fiber. By "localized mode" we mean a focused spot at a given position on the fiber end face. At the other end of the fiber, the electric field emerging from the fiber is measured on a camera by off-axis holography and each pixel of the camera is considered a localized output mode. Input and output polarizations are controlled by waveplates just before and after the fiber, respectively. 
A simplified flowchart of the measurement and subsequent data treatment is presented in Fig.~\ref{fig:TMmeas}(b).
First, a stack of images of the output electric field is measured for every input localized mode (in practice 4 stacks, one for each combination of input and output polarization). This stack is re-organized into a two-dimensional complex matrix in input localized modes (rows) and output localized modes (columns)---this is the "system TM" in the basis of localized modes. This system TM contains a contribution from the fiber itself (the fiber TM) as well as "misalignment" contributions from the free-space optics between SLM and fiber; and between fiber and camera---we will seek to extract the fiber TM from this system TM. To this end we perform a singular value decomposition of the system TM and retain only the 10 first singular vectors (the first 10, corresponding to the number of guided modes, have singular values significantly larger than the rest). We then run an iterative optimization algorithm that determines the misalignment contribution: in each iteration the algorithm makes a guess for the misalignment, modifies the basis of calculated modes by this misalignment, and then compares the mode space spanned by the modified calculated modes to the mode space spanned by the retained singular vectors, the algorithm continues to iterate and optimize its guess for the misalignment until it converges on maximum likeness between the two mode spaces. In a final step, the misalignment is numerically removed from the system TM, and a basis change is performed to express it in the basis of the calculated fiber modes. This is the TM of the fiber alone, the fiber TM, and we insist on the fact that our method is capable of isolating the contribution of the fiber from misalignment contributions like slow mechanical drift which can lead to contributions that would otherwise be interpreted as stemming from the fiber.\\
With the current setup the measurement of the entire TM takes around 40 minutes when a basis of 25x25 localized modes is used. We measured a stack of fiber TMs over a time span of 36~h with the smallest time interval between subsequent measurements being 2~h. 
The measurements were performed on an unconstrained fiber lying on an optical table in a half-turn shape in a laboratory room equipped with a standard air-condition unit, but no other special environmental control measures were taken. The temperature in vicinity (few cm) of the fiber was logged every second. 

\section{\label{sec:Results}Results}
\subsection{Temporal evolution of the fiber TM}
\begin{figure}[htbp]
\centering
\includegraphics{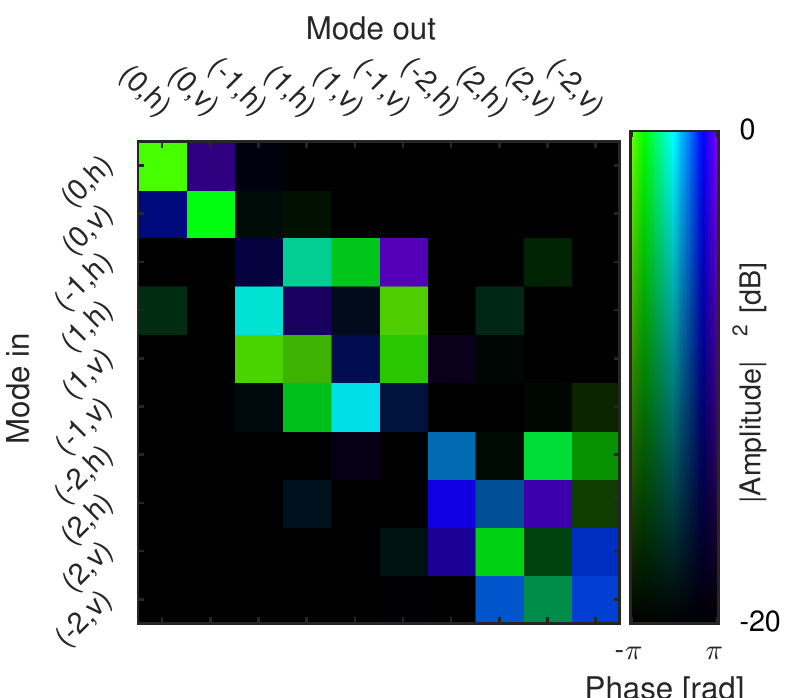}
\caption{
\label{fig:lin_TMex}
One example extracted from the stack of 11 measured fiber TMs. Notation of the modes as in Tab.~\ref{tab:LP}. Note the doubly-graduated color map where color codes for the phase and the saturation codes for the norm-squared amplitude of the complex-valued fiber TM. 
}
\end{figure}
\begin{figure}[htbp]
\centering
\includegraphics{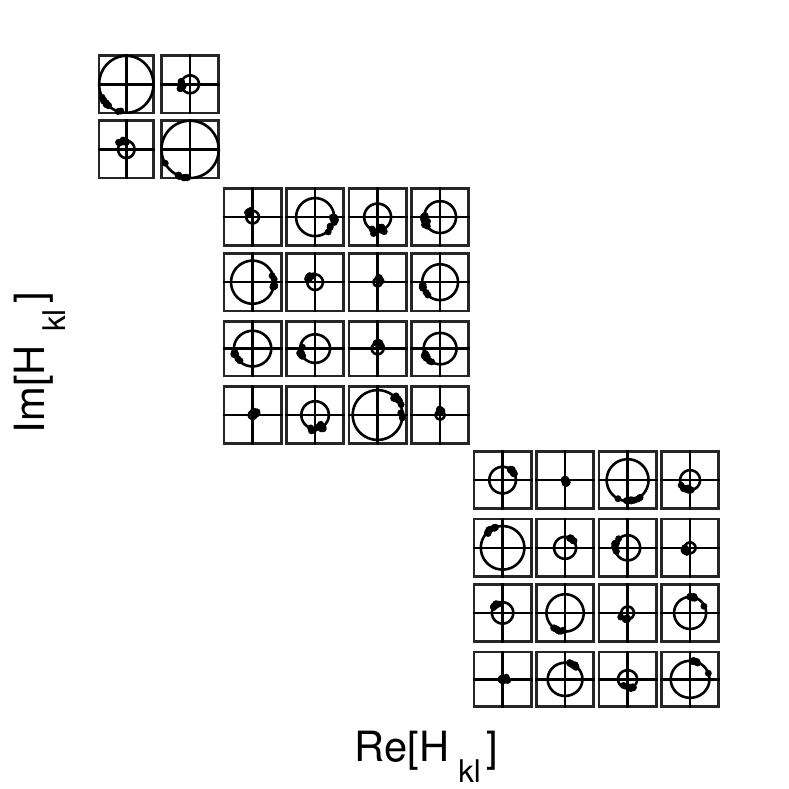}
\caption{
\label{fig:lin_cplx}
Representation in the complex plane of the fiber TM elements measured at different times. Horizontal and vertical lines represent the real and imaginary axes. Depicted circles are guides to the eye and have radius equal to the time-averaged norm of the fiber TM element. 
}
\end{figure}

We acquired a stack of 11 fiber TMs over a time span of 36~h. The stack is indexed as follows
\begin{equation}
H_{kl}(t_{i})
\end{equation}
with $k$ the index of the input mode, $l$ the index of the output mode, and $i$ the temporal index.
Figure~\ref{fig:lin_TMex} shows one of these complex-valued fiber TMs. As expected, the matrix elements of significant amplitude are found in three blocks of size 2$\times$2, 4$\times$4, and 4$\times$4 on the diagonal constituted by the three mode families with $|l|$~=~0, 1, and 2 (cf Tab.~\ref{tab:LP}), or in terms of the vector modes (HE$_{11}^{(e)}$, HE$_{11}^{(o)}$); (TE$_{01}$, HE$_{21}^{(e)}$, HE$_{21}^{(o)}$, TM$_{01}$); and (HE$_{31}^{(e)}$, HE$_{31}^{(o)}$, EH$_{11}^{(e)}$, EH$_{11}^{(o)}$). The small effective index differences within a family allow its members to couple freely in the presence of small perturbations of the fiber, while the large effective index differences between mode families ($>10^{-3}$ cf Tab.~\ref{tab:fiber}) impede coupling (For reference, polarization-maintaining fibers typically have $10^{-4}$ index difference between the fast and slow axes). 
\begin{figure}[htbp]
\centering
\includegraphics{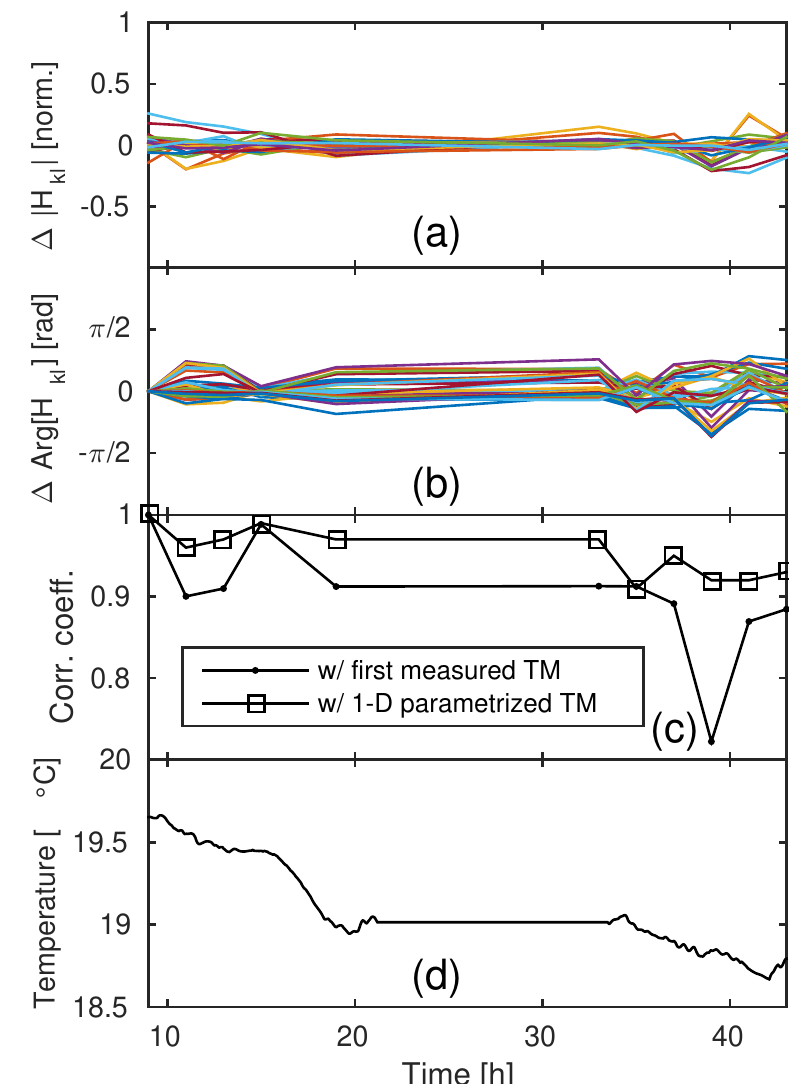}
\caption{
\label{fig:lin_TMevol}
Temporal evolution of the fiber TM. (a) Normalized amplitude variation of fiber TM elements. (b) Phase variation of TM elements. Curves for fiber TM elements with negligible amplitude ($<0.18$) are not shown. (c) Correlation coefficient between (dots) the first measured fiber TM and the ones measured at time $t_{i}$; (squares) the TM parametrized in one parameter and the fiber TM measured at time $t_{i}$. (d) Log of the temperature in the vicinity of the fiber during the measurement window.
}
\end{figure}
We start the investigation of the temporal evolution of the fiber TM by examining the amplitudes and the phases of the elements of the fiber TMs. Figure~\ref{fig:lin_TMevol}(a) presents the temporal variation of the normalized amplitude defined as
\begin{equation}
\Delta|H_{kl}|(t_{i}) = \frac{|H_{kl}(t_{i})| - \langle | H_{kl}(t_{i}) | \rangle _{i}}{\langle | H_{kl}(t_{i}) | \rangle _{i}}, 
\end{equation}
which can take on values in the interval $[0;1]$. $\langle \cdots \rangle_{i}$ denotes the average over the time index $i$. 
And Fig.~\ref{fig:lin_TMevol}(b) presents the temporal variation of the phase defined as
\begin{equation}
\Delta \mathrm{Arg} [ H_{kl}(t_{i}) ] = \mathrm{Arg} [ H_{kl}(t_{i}) ] - \mathrm{Arg} [ H_{kl}(t_{0}) ]
\end{equation}
which can take on values in the interval $[-\pi;\pi[$. 
In Figs.~\ref{fig:lin_TMevol}(a) and \ref{fig:lin_TMevol}(b) only curves for those fiber TM elements whose mean amplitude $>0.18$ are shown (27 out of 36 elements contained in the diagonal blocks), because of the higher relative noise on the remaining elements. 
It can be appreciated that the variations are quite small, the amplitude varying only within 25~$\%$ of the mean and the phase only within $\pi/2$. From the fact that there are no large phase jumps we conclude that the time resolution of our fiber TM measurement (40~min) is sufficient to resolve the slow drift under our experimental conditions. \\
We further examine the time evolution of the fiber TM by computing the correlation coefficient between the first measured fiber TM at time $t_{0}$ with those measured at later times $t_{i}$. We define the correlation coefficient as
\begin{equation}
C_{0i} = \frac{ \sum_{k,l} H_{kl}^{*}(t_{0}) H_{kl}(t_{i})}{ \sqrt{\sum_{k,l}|H_{kl}(t_{0})|^{2}} \sqrt{\sum_{k,l}|H_{kl}(t_{i})|^{2}} }. 
\end{equation}
where $^{*}$ denotes element-wise complex conjugation. The evolution of this correlation coefficient is presented in Fig.~\ref{fig:lin_TMevol}(c) (dots). The correlation coefficient remains above 70~$\%$ throughout which, again attests to the slow drift of the fiber TM. \\
For reference, the temperature in close vicinity of the fiber was monitored during the measurements, it is shown in Fig.~\ref{fig:lin_TMevol}(d). \\
An equivalent, but perhaps slightly more illustrative, representation of the same dataset is presented in Fig.~\ref{fig:lin_cplx}. For each element, all 11 complex-valued fiber TM elements acquired at times $t_{i}$ are plotted in the complex plane. Horizontal and vertical lines represent the real and imaginary axes, and the circles are guides to the eye with radius equal to the absolute value of the matrix element. This illustration allows us to observe that the drift in the fiber TM can be attributed mainly to a drift in the phase since all the 11 plotted points per sub-figure overwhelmingly tend to lie on the circles. This representation does not however provide any insight on the correlation between the phase variations of the different elements, which is the topic of the next paragraph.

\subsection{Parametrization of the TM}
\begin{figure}
\centering
\includegraphics{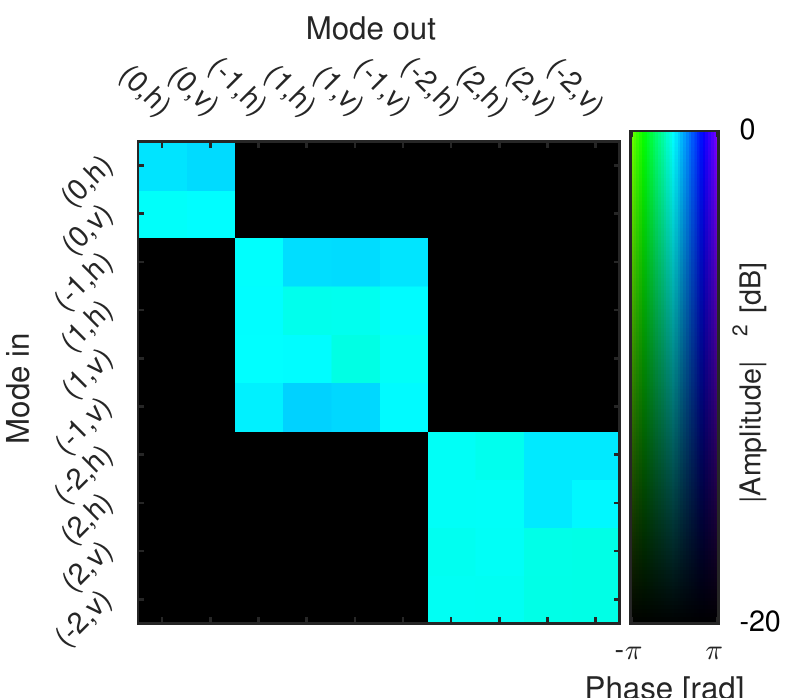}
\caption{\label{fig:lin_SV}
First singular vector of $\Delta \Phi$ in complex form---drift operator in the case of parametrization in one parameter. 
}
\end{figure}

We will attempt to find a parametrization of the fiber TM that describes its temporal evolution. 
We start out from the stack of fiber TMs $H_{kl}(t_{i})$ and the observation made in the previous paragraph that mainly the phase variation of the fiber TM needs to be taken into account. 
We create the two-dimensional matrix $\Delta \Phi_{\{kl\}i}$ in the composite index $\{kl\}$ and the temporal index $i$:
\begin{equation}
\Delta \Phi _{ \{kl\}i} = \mathrm{Arg}[ H_{kl}(t_{i}) ]- \mathrm{Arg}[ H_{kl}(t_{0})], (k,l) \in \mathcal{D}
\end{equation}
where $\mathcal{D}$ is the group of indices ($k$,$l$) of the diagonal blocks of the fiber TM. The matrix has dimension 36 x 11 (number of elements in the diagonal blocks x number of time points). We take the singular value decomposition of the resulting matrix (we omit the indices for brevity):
\begin{equation}
\Delta \Phi = U S V ^{\dagger}. 
\end{equation}
The first matrix U contains the left-singular vectors, S is a diagonal matrix formed by the
singular values and V contains the right-singular vectors.
We find a first singular value which is significantly larger than the rest. The first singular value accounts for 35~$\%$ of the singular weight; the first three for 69~$\%$, and the first five for 88~$\%$. 
For now, we retain only the first singular vector which amounts to taking the first column of $U$, $U_{\{kl\}1}$, with which we can construct an empirical, parametrized version of the fiber TM (in the parameter $\alpha$):
\begin{equation}
H_{kl}^{\mathrm{(param)}}(\alpha) =  \Bigg \{ \begin{array}{ll}
H_{kl}(t_{0}) \cdot \mathrm{exp} ( {j  \alpha U_{\{kl\}1}}) & , (k,l) \in \mathcal{D} \\
H_{kl}(t_{0}) & , (k,l) \notin \mathcal{D}
\end{array} 
\end{equation}
The drift is thus mathematically expressed as a complex-valued "drift" operator, identifiable as the matrix being multiplied onto $H_{kl}(t_{0})$ in the Equation above. In Fig.~\ref{fig:lin_SV} this drift operator is presented.
We now check whether the parametrized TM is a good model of the multi-modal channel represented by the fiber. To do so, for each fiber TM of the fiber TM stack we identify the $\alpha$ that results in the largest inner product
\begin{equation}
C^{\mathrm{(param)}}_{i} = \frac{ \sum_{k,l} [H^{\mathrm{(param)}}_{kl}(\alpha)]^{*} H_{kl}(t_{i})}{ \sqrt{\sum_{k,l} |H^{\mathrm{(param)}}_{kl}(\alpha)|^{2}} \sqrt{\sum_{k,l} |H_{kl}(t_{i})|^{2}  }}. 
\end{equation}
The results are summarized in Tab.~\ref{tab:lin_parametrization}. It can be appreciated that this simple parametrization in only a single parameter gives a very decent description of the measured TMs with its 36 significant elements, with $C_{i}^{\mathrm{(param)}}$ systematically above 0.9. The same values are also shown in Fig.~\ref{fig:lin_TMevol}(c) as squares. \\
\begin{table}[htbp]
\caption{\label{tab:lin_parametrization} Correlation coefficients with the TM parametrized in one parameter. }
\begin{displaymath}
\begin{array}{|c|c|c|}
  \hline
  \mathrm{Time [h]} & \alpha & C_{i}^{\mathrm{(param)}} \\
  \hline
  9 & 0.00 & 1.00 \\  
  \hline
  11 & 1.54 & 0.96 \\
  \hline
  13 & 1.49 & 0.97 \\
  \hline
  15 & 0.17 & 0.99 \\  
  \hline
  19 & 1.48 & 0.97 \\
  \hline
  33 & 1.48 & 0.97 \\
  \hline
  35 & -0.15 & 0.91 \\
  \hline
  37 & 1.61 & 0.95 \\
  \hline
  39 & 3.04 & 0.92 \\
  \hline
  41 & 1.43 & 0.92 \\
  \hline
  43 & 1.41 & 0.93 \\
  \hline
\end{array}
\end{displaymath}
\end{table}
It is clear that parametrizing the TM in more parameters will result in ever better concordance with the measured fiber TMs. Nevertheless, we can conclude from our results that---in our measurement conditions---drift results in changes to the fiber TM that are highly correlated between matrix elements. Thus, even a low-dimensional parametrization with number of parameters much smaller than the number of matrix elements involved can describe the measurement very well. 

\section{\label{sec:Discussion}Discussion}
The appeal of measuring the full fiber TM is that it is a direct measurement of the linear behavior of the fiber. It thus contains more information than indirect measurements like time-averaged cross-talk and bit-error ratio. As such, fiber TM measurements could be a good basis for developing empirical multi-modal channel models but could also be a good basis for comparison with various multi-modal channel models that have been developed \cite{AntonelliOE2013, HoOE2011}. A fiber TM measurement resolved in wavelength, as was done in \textit{e.g.} Ref.~\citenum{CarpenterNatPhoton2015} would be required to elucidate also the group-delay effects for comparisons with models of group-delay behavior \cite{HoJLT2011, HoIEEEPTL2012}. \\
The length of the fiber studied here was very short, 1.5~m, but it is of the same length scale as some of the multi-mode components which would be required for potential future multi-modal fiber optic communication networks, like mode filters or multi-mode amplifiers \cite{TrinelOE2016, TrinelOFT2016}. So our method could have an appeal for characterizing this type of multi-mode components. Fiber lengths that are pertinent for data transmission would of course be orders of magnitude longer. As the time scale of drift likely decreases as a function of length the fiber TM measurement would have to be made faster to retain its relevance for characterizing transmission fibers. The current time to measure a fiber TM, defining the temporal resolution, is 40 minutes. In principle this could be sped up in a number of ways that we have not yet implemented. We currently measure the four combinations of input and output polarizations separately, by measuring them simultaneously in a two-polarization setup an immediate gain of four would be achieved and, in addition one would eliminate the source of error which is phase drift between measurement of different polarization combinations. Another way would be to reduce the number of local modes used in the initial measurement from the current 25x25 which is massively over-sampling the mode space of the fiber which guides only 10 vector modes; this would yield a proportional gain in measurement time. Finally, one could envision employing a SLM with shorter response time. Alternatively, we could measure directly in the basis of fiber modes, but this would preclude the ability to isolate the fiber TM from the misalignment. \\

\section{\label{sec:Conclusion}Conclusion}
We have presented a method for measuring the fiber TM of a ring-core FMF without misalignment contributions from the interfacing optics, and we have shown how a simple, low-dimensional parametrization of the fiber TM of a short length of FMF gives a good description of the temporal evolution of the multi-modal channel it represents over a time span of two days. 
We believe that our approach could aid in developing accurate channel models for few-mode fiber optic transmission systems as well as a deeper understanding of FMFs. 

\begin{acknowledgments}
This work has been partially supported by the Agence Nationale de la Recherche through the LABEX CEMPI (ANR-11-LABX-0007) and the Equipex Flux (ANR-11-EQPX-0017), as well as by the Ministry of Higher Education and Research, Hauts de France council and European Regional Development Fund (ERDF) through the Contrat de Projets Etat-Region (CPER Photonics for Society P4S). FUI MODAL (FUI-AAP19). 
\end{acknowledgments}

\appendix

\section{\label{app:TM}Detailed description of the transmission matrix measurement}
Here we describe in detail the measurement of the fiber TM at time $t_{i}$, a measurement which is repeated at several times to produce the stack of fiber TMs presented in the main manuscript. 
The experimental setup is the one presented in Fig.~\ref{fig:TMmeas}(a). 
We use a coherent laser source (Yenista Tunics T100S) at wavelength of 1550~nm. A polarization-maintaining fiber optic coupler (not shown) divides the laser into two different beams: the signal beam and the reference beam. The signal beam is collimated, passes through a polarizing beam splitter (PBS) and reaches the screen of a two-dimensional spatial light modulator (SLM, Meadowlark P1920-1625-HDMI). The polarization axis of the signal fiber, the PBS, and the SLM are parallel. We use the SLM screen as a diffraction grating by displaying a saw-tooth phase mask with desired horizontal and vertical periodicity, so we can change the reflection angle of the beam and localize the beam on a defined spot: we scan the fiber face with our beam spot so we cover all the possible injection positions in our fiber. Each angle is considered as a localized input mode. Between the SLM and the fiber, we use a half-wave plate (HWP) to generate the desired polarization, either horizontal or vertical. The beam emerges from the tested fiber, is collimated, passes through another HWP and a PBS. The HWP will convert one polarization component of the emerging beam into a linear polarization, parallel to the axis of the PBS. Then, the signal beam reaches a non-polarizing beam splitter (BS). On the other output of the fiber optic coupler, the reference beam travels approximately an equal optical path as the signal beam, while passing through a HWP and a PBS which are aligned in the same direction as the HWP and PBS of the emerging signal beam. A mirror is used to guide the reference beam into the BS, where it joins the signal beam. The two beams end their path at the sensor of an infra-red camera (CAM), with an angle slightly higher than 0 degrees, enabling off-axis holographic recording. On the camera, an interference image appears as a result of the interference between signal and reference beams. An interference image is saved for each localized input mode, and for each combination of input and output polarization. The images form four stacks, one stack for each polarization combination. Four acquisitions are needed with different angles between the axes of the HWP and polarization axis of the SLM: the first acquisition for the input and output HWP at 0 degrees, a second acquisition with 0 degrees for the input HWP and 45 degrees for the output HWP, the third acquisition with 45 degrees at the input and 0 at the output, and the last one with both HWPs at 45 degrees. In this example, we chose to operate with a resolution of 25$\times$25 localized input modes, and therefore each stack contains 625 images. 

\begin{figure}
	\centering
	\includegraphics[width=\columnwidth]{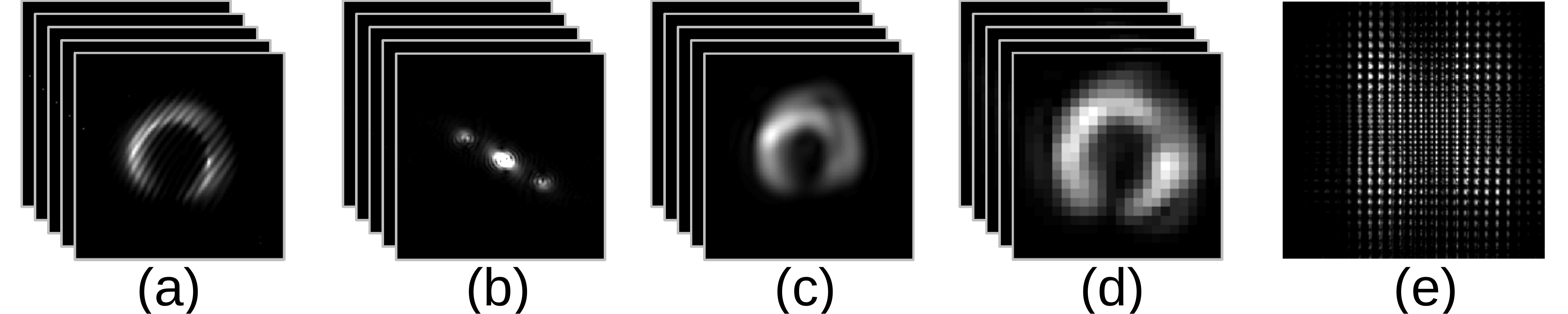}
	\caption{\label{fig:app1}
	(a) Stack of interference images for 1 polarization combination. (b) Fourier transformation of the stack. (c) Inverse Fourier transformation of the filtered stack. (d) Down sampling of the images. (e) The transmission matrix of the stack. Only the norm of the complex-valued matrices is shown. 
	 	}
\end{figure}

\begin{figure}
	\centering
	\includegraphics[width=\columnwidth]{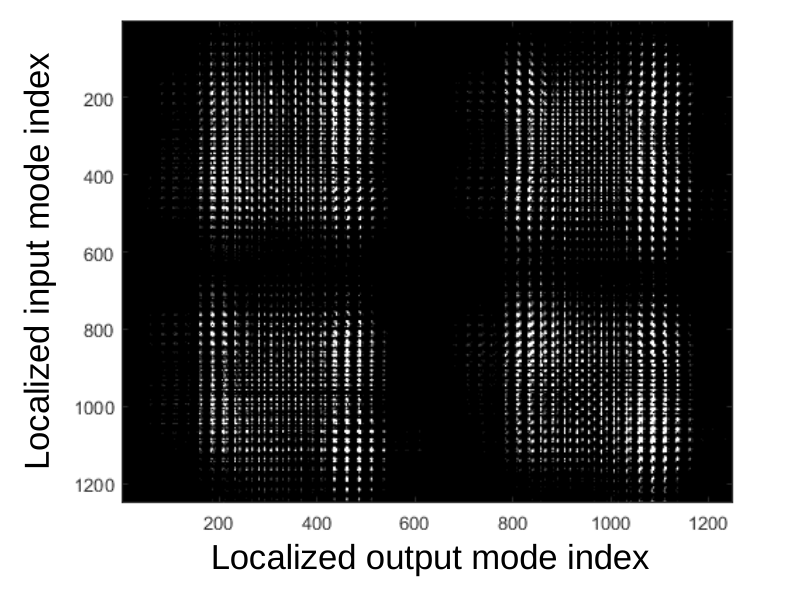}
	\caption{\label{fig:app2}
	The total system TM. Only the norm of the complex-valued matrix is displayed. 
	}
\end{figure}

Once the four stacks acquired we begin the data analysis: each stack is analyzed on its own in the first step [Fig.~\ref{fig:app1}(a)]. First, a two-dimensional Fourier transformation is applied on each image of the stack. The result is a stack of images with 3 peaks in the Fourier domain [Fig.\ref{fig:app1}(b)]. We then apply a filter on the images to isolate one first-order peak and translate it to zero spatial frequency. An inverse Fourier transformation is then applied to give a stack of images of the electric field on CAM [Fig.~\ref{fig:app1}(c)]. We then down sample the images to reduce the calculation time, while retaining of the information on the output electric field, from 320$\times$256 the resolution of CAM to 25$\times$25, matching the number of elements in an image to the number of images in the stack [Fig.~\ref{fig:app1}(d)]. This stack of images is reshaped into a two-dimensional complex matrix
	that will contain all the input localized modes in its rows, and all the output localized modes in its columns. This matrix is one quadrant of the system TM (one polarization combination out of four possible)  [Fig.~\ref{fig:app1}(e)] containing contribution from the fiber itself as well as all the other optical components between the SLM and CAM, with all their misalignment. This matrix is expressed in the basis of the localized modes and has a 625$\times$625 elements.
The 4 quadrants of the system TM are combined to form the total system TM that links all the input and output localized modes and their polarization states, Fig.~\ref{fig:app2}. This system TM has dimensions 1250$\times$1250 and is denoted $T_{kl}$. 

\begin{figure}[htbp]
	\centering
	\includegraphics[width=\columnwidth]{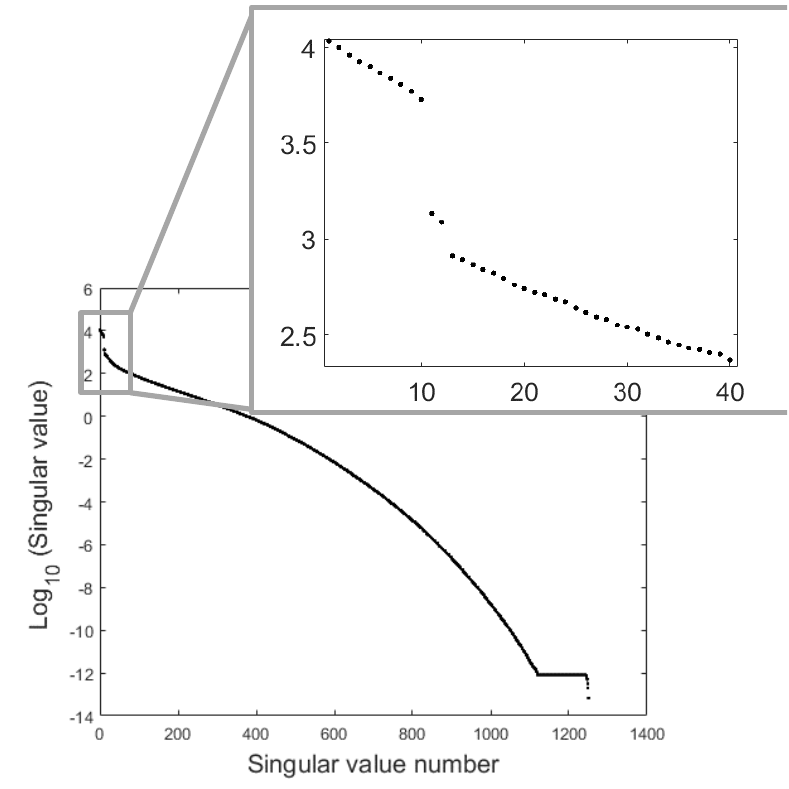}
	\caption{\label{fig:app3}
	The singular values found in the singular value decomposition of the system TM. Logarithmic scale.  
	}
\end{figure}

As a first step in the data treatment which will factor the system TM into fiber TM and misalignment operators, we perform a singular value decomposition on the system TM (omitting the indices for brevity), 
\begin{equation}
T = USV^{\dagger}. 
\end{equation}
The first matrix $U$ contains the left-singular vectors of the input basis, S is a diagonal matrix formed by the singular values and $V$ contains the right-singular vectors. The matrix $S$ will give us indications about the number of guided modes: a significant singular value mark the presence of a guided mode. In our current example, we have 10 significant singular values cf Fig.~\ref{fig:app3}, hence 10 guided modes. The 10 first singular vectors contained in the 10 first columns of the two matrices $U$ and $V$ are used to define the modal content of the fiber. 
\begin{figure}[htbp]
\centering
\includegraphics[width=\columnwidth]{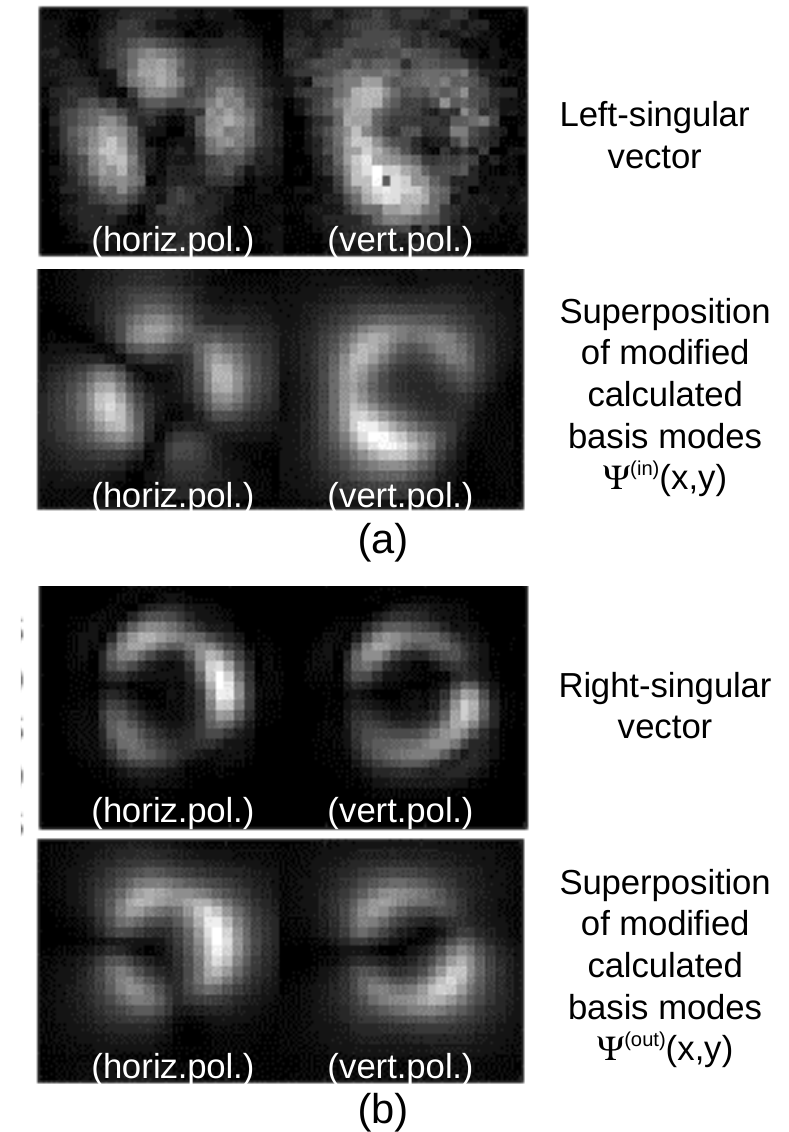}.
\caption{\label{fig:app4}
Comparison of (a) a left-singular vector and (b) a right-singular vector extracted from the singular-value decomposition of the system TM with a linear combination of the modified calculated basis modes found by iterative optimization algorithm. 
Only the norm of the fields are shown. 
}
\end{figure}
\begin{figure}[htbp]
\centering
\includegraphics{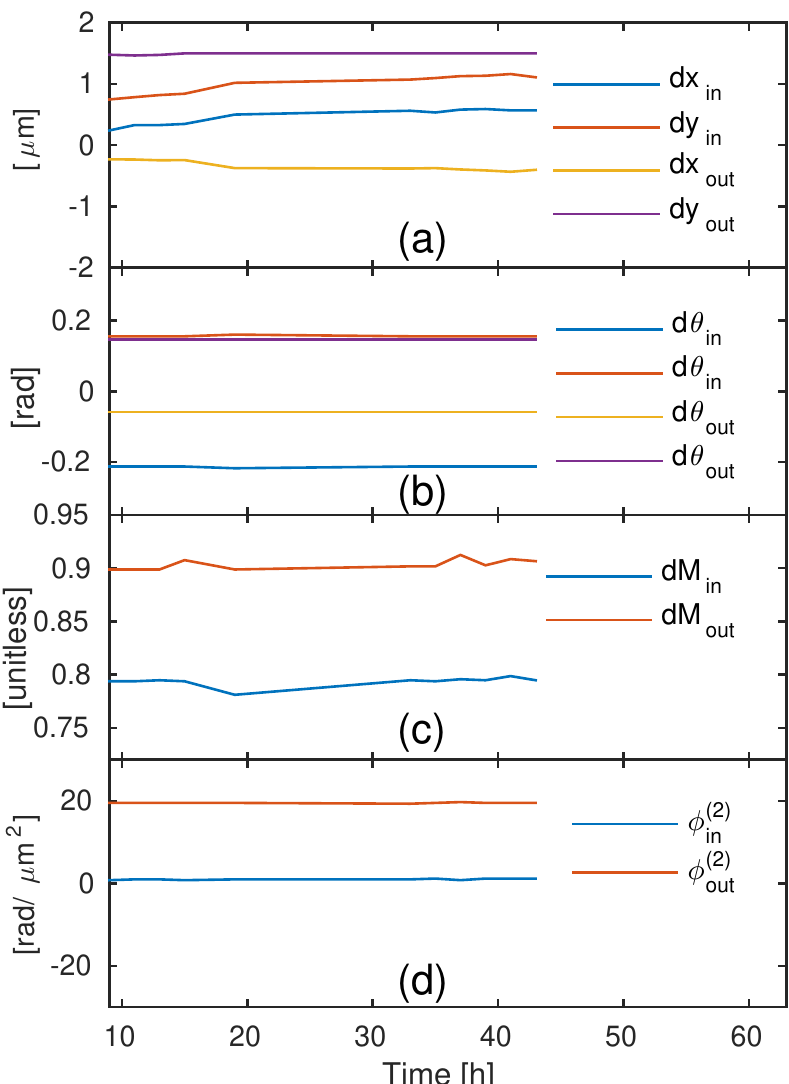}
\caption{\label{fig:app5}
Time evolution of the misalignment parameters. 
}
\end{figure}

An iterative optimization algorithm now makes a guess for the misalignment parameters ($dx_{\mathrm{in}}$, $dy_{\mathrm{in}}$, $d\theta_{x,\mathrm{in}}$, $d\theta_{y,\mathrm{in}}$, $\phi_{\mathrm{in}}^{(2)}$, $dM_{\mathrm{in}}$, $dx_{\mathrm{out}}$, $dy_{\mathrm{out}}$, $d\theta_{x,\mathrm{out}}$, $d\theta_{y,\mathrm{out}}$, $\phi_{\mathrm{out}}^{(2)}$, $dM_{\mathrm{out}}$). 
From the basis of calculated mode fields $\Psi_{k}(x,y)$ (Tab.~\ref{tab:LP}) it generates a basis of modified calculated input and output mode fields $\Psi_{k}^{(\mathrm{in})}(x,y)$ and $\Psi_{l}^{(\mathrm{out})}(x,y)$ by applying the "misalignement operators" $M^{(\mathrm{in})}(x,y)$ and $M^{(\mathrm{out})}(x,y)$, 
\begin{eqnarray}
\Psi_{k}^{(\mathrm{in})}(x,y) &=& M^{(\mathrm{in})}(x,y) \Psi_{k}(x,y) \\
\Psi_{l}^{(\mathrm{out})}(x,y) &=& M^{(\mathrm{out})}(x,y) \Psi_{l}(x,y)
\end{eqnarray}
where
\begin{eqnarray}
  &&M^{(\mathrm{in})}(x,y) = \mathrm{exp} \left( j \phi_{in}^{(2)} (x^{2} + y^{2}) \right) \nonumber \\
  && \times \mathcal{F}^{-1} \mathrm{exp} \left( j 2\pi f_{x} dx_{in} + j 2\pi f_{y} dy_{in} \right)  \mathcal{F} \nonumber \\
  && \times \mathrm{exp} \left( \frac{j 2\pi}{\lambda} x \mathrm{sin} (d\theta_{x,in}) +  \frac{j 2\pi}{\lambda} y \mathrm{sin} (d\theta_{y,in}) \right) \nonumber \\
\end{eqnarray}
and 
\begin{eqnarray}
  &&M^{(\mathrm{out})}(x,y) = \mathrm{exp} \left( j \phi_{out}^{(2)} (x^{2} + y^{2}) \right) \nonumber \\
  && \times \mathcal{F}^{-1} \mathrm{exp} \left( j 2\pi f_{x} dx_{out} + j 2\pi f_{y} dy_{out} \right) \mathcal{F} \nonumber \\
  && \times  \mathrm{exp} \left( \frac{j 2\pi}{\lambda} x \mathrm{sin} (d\theta_{x,out}) +  \frac{j 2\pi}{\lambda} y \mathrm{sin} (d\theta_{y,out}) \right)  \nonumber \\
\end{eqnarray}
where $\mathcal{F}$ and $\mathcal{F}^{-1}$ denote forward and inverse Fourier transform and $f_{x}$ and $f_{y}$ the spatial frequencies. The parameters $dM_{\mathrm{in}}$ and $dM_{\mathrm{out}}$ are left out of the misalignment operators, but in the algorithm they are basically used as scaling parameters, amounting to $(x,y) \rightarrow (dM_{\mathrm{in}} x, dM_{\mathrm{in}} y)$ and $(x,y) \rightarrow (dM_{\mathrm{out}} x, dM_{\mathrm{out}} y)$

The algorithm calculates the inner products between the modified calculated basis and the singular vectors:
\begin{eqnarray}
 C^{(\mathrm{in})}_{km}&=& \langle \Psi_{k}^{(\mathrm{in})}(x,y) | U_{m}(x,y) \rangle  \\
 C^{(\mathrm{out})}_{lm}&=& \langle \Psi_{l}^{(\mathrm{out})}(x,y) | V_{m}(x,y) \rangle 
\end{eqnarray}
and calculates the penalty functions
\begin{eqnarray}
P_{\mathrm{in}} &=& 1 - \frac{1}{N}\sum_{km} |C_{km}^{(\mathrm{in})}|^{2} \\
P_{\mathrm{out}} &=& 1 - \frac{1}{N}\sum_{lm} |C_{lm}^{(\mathrm{out})}|^{2} \\
\end{eqnarray}
with $N$~=~10 the number of modes guided by the fiber. $P_{\mathrm{in}}$ ($P_{\mathrm{out}}$) takes on values in the interval $[0;1]$, is equal to zero for perfect likeness between the calculated input (output) mode space and the mode space of the retained left-singular (right-singular) vectors, and equal to one for no likeness. 
The algorithm iterates, refining its guess for the misalignment parameters, until it converges on a minimum in $P_{\mathrm{in}}$ and $P_{\mathrm{out}}$. The optimizations of input and output parameters are done separately. 
Finally, the fiber TM is constructed as
\begin{equation}
H_{kl} = C^{(\mathrm{in})}_{km} (C^{(\mathrm{out})}_{lm})^{\dagger}, 
\end{equation}
giving a matrix like Fig.~\ref{fig:lin_TMex}. \\
A visual gauge of the quality of the solution found by the algorithm can be obtained by comparing a left-singular (right-singular) vector to the closest semblence of it that can be constructed by a linear combination of the $\Psi^{(\mathrm{in})}(x,y)$ ($\Psi^{(\mathrm{out})}(x,y)$) onto which the algorithm has converged. Such a comparison is given in Fig.~\ref{fig:app4}.

Finally, the misalignment parameters returned by the algorithm are presented in Fig.~\ref{fig:app5}.

\section{\label{app:altbasis} Measured fiber TMs expressed in other mode bases}
\begin{figure}[htbp]
\centering
\includegraphics{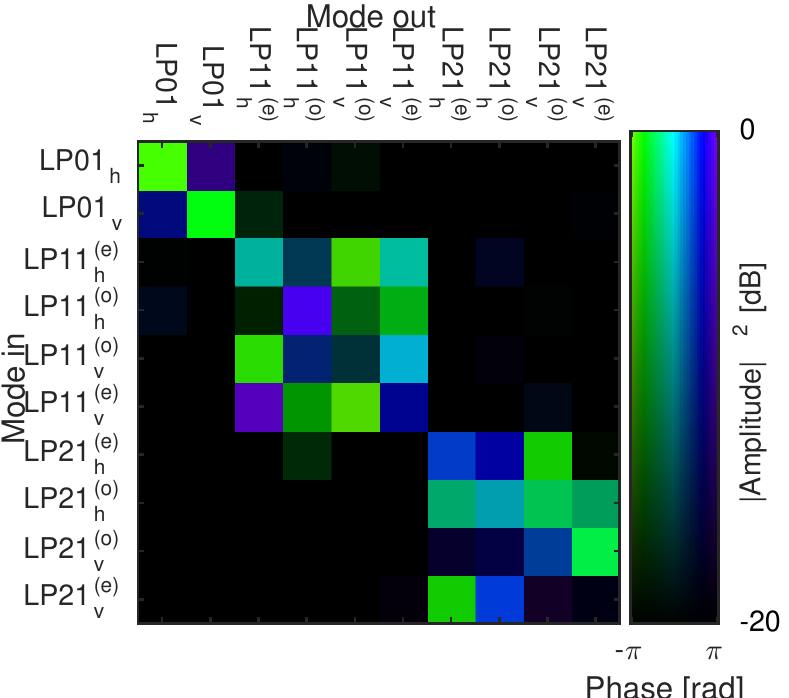}
\caption{\label{fig:app6a} 
The fiber TM of Fig.~\ref{fig:lin_TMex} expressed in the basis of the standard linearly-polarized (LP) modes. 
}
\end{figure}

\begin{figure}[htbp]
\centering
\includegraphics{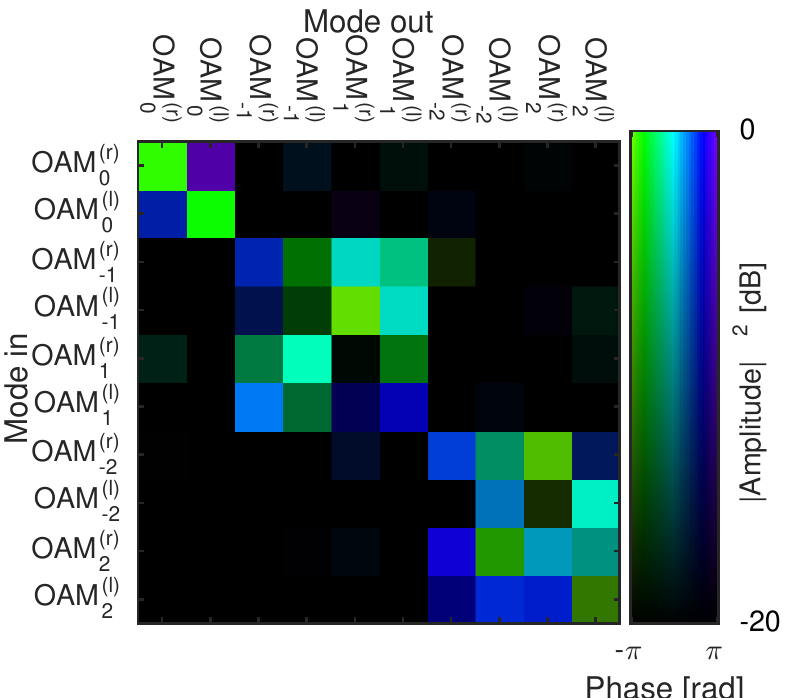}
\caption{\label{fig:app6b}
The fiber TM of Fig.~\ref{fig:lin_TMex} expressed in the basis of the standard orbital angular momentum (OAM) modes. 
}
\end{figure}

The fiber TM presented in the main text can be transformed to any modal basis by applying a basis change matrix to it. Here we present the fiber TM of Fig.~\ref{fig:lin_TMex} expressed in two commonly-used modal bases, the basis of linearly-polarized (LP) modes (Fig.~\ref{fig:app6a}) and the basis of orbital angular momentum (OAM) modes (Fig.~\ref{fig:app6b}). 
As can be seen the detailed appearance of the fiber TM changes slightly under basis transformation, but its overall appearance with three diagonal blocks of size  2$\times$2, 4$\times$4, and 4$\times$4 remains. 

\nocite{*}

%

\end{document}